\documentclass[fleqn,9pt,lineno,twocolumn]{article}
\usepackage{amsmath}
\usepackage[english]{babel}
\usepackage[letterpaper,top=2cm,bottom=2cm,left=2cm,right=2cm,marginparwidth=1.75cm]{geometry}
\usepackage{braket}
\usepackage{algorithm}
\usepackage[noend]{algpseudocode}
\usepackage{float}
\usepackage{graphicx}
\usepackage{hyperref}
\usepackage{authblk}
\usepackage{afterpage}

\begin{document}
%author[1]{First Author}
%\author[2]{Second Author}
%\affil[1]{Address of first author}
%\affil[2]{Address of second author}

\title{Broad Yet Narrow: Super-resolution techniques to simulate electronic spectra of large molecular systems}
%\author{Matthias Kick, Ezra Alexander, Anton Beiersdorfer and Troy Van Voorhis}
\author[$\dagger$]{Matthias Kick\footnote{mkick@mit.edu}}
\author[$\dagger$]{Ezra Alexander}
\author[$\ddag$]{Anton Beiersdorfer}
\author[$\dagger$]{Troy Van Voorhis}
\affil[$\dagger$]{Department of Chemistry, Massachusetts Institute of Technology, Cambridge, Massachusetts 02139, USA}
\affil[$\ddag$]{Technical University of Munich, Lichtenbergstrasse 4, Garching 85747, Germany}

\maketitle
\begin{abstract}
% original abstract 172 words
An accurate treatment of electronic spectra in large systems with a technique such as time dependent density functional theory (TDDFT) is computationally challenging. Due to the Nyquist sampling theorem, direct real time simulations must be prohibitively long in order to recover a suitably sharp resolution in frequency space. Super-resolution techniques such as compressed sensing  and MUSIC assume only a small number of excitations contribute to the spectrum, which fails in large molecular systems where the number of excitations is typically very large. We present a new approach that combines exact short time dynamics with approximate frequency space methods to capture large narrow features embedded in a dense manifold of smaller nearby peaks. We show that our approach can accurately capture narrow features and broad quasi-continuum of states at the same time - even when the features overlap in frequency. Our approach is able reduce the required simulation time by a factor of 20-40 with respect to standard Fourier analysis and shows promise for the accurate whole-spectrum prediction of large molecules and materials.
% shortend abstract 149 words
%The Nyquist theorem makes an accurate treatment of electronic spectra in large systems with a technique such as time dependent density functional theory (TDDFT) computationally challenging. Real time simulations must be prohibitively long in order to achieve sufficient resolution in frequency space. Super-resolution techniques such as compressed sensing  and MUSIC assume only a small number of excitations contribute to the spectrum, which fails in large molecular systems where the number of excitations is typically very large. We present a new approach that combines exact short time dynamics with approximate frequency space methods. Our approach can accurately capture narrow features and broad quasi-continuum of states at the same time - even when the features overlap in frequency. The method is able reduce the required simulation time by a factor of 20-40 with respect to standard Fourier analysis and shows promise for the accurate whole-spectrum prediction of large molecules and materials.

\end{abstract}
\section*{Introduction}
Electronic excitations in molecules and materials are important for understanding various kind of phenomena such as photo-excitation in solar cells, optical excitations in OLEDS and quantum dots(QDs).\cite{REN2021113087,Goldzak2021,D0CP00060D,Neef2023,SLOWIK2017132,Kinoshita2015,Gasparini2021,COPPOLA2022125603,10.1093/oxfmat/itad003,ZAIER2019108,D2RA06880J} Theoretically, electronic excitations can be obtained by analysing the frequency components of the time-dependent dipole moment obtained from real-time propagation.\cite{ptddft} Among various other methods such GW/BSE\cite{PhysRevLett.128.016801}, EOM\cite{doi:10.1021/acs.jctc.0c00639,10.1063/5.0004865,10.1063/5.0099192} or ADC\cite{C7CP07849H}, time-dependent density functional theory (RT-TDDFT)\cite{ptddft} is the most promising method to calculate the whole spectrum of large systems due to its superior scaling with respect to system size compared to other methods. Because of the computational complexity of real time simulations for large molecules and materials, one is typically restricted to fairly short time dynamics (e.g. tens of fs). Due to the Nyquist sampling theorem, discrete Fourier analysis of the short time dynamics fails to capture the narrow features that are critical fingerprints of molecular spectra.
Meanwhile, standard super-resolution methods - such as compressed sensing (CS) \cite{compressed_sensing,SEJDIC201822,Orovic2016}, MUSIC\cite{music,1143830} and orthogonal matched pursuit \cite{omp,258082} - typically fail for large molecular systems because they require the number of narrow features to be small, whereas the spectra of large molecules tends to be quite densely populated. Similarly, linear response approaches like the Casida \cite{casida_annual} or Sternheimer equation \cite{PhysRev.84.244} typically require one-at-a-time identification of roots and likewise fail when the number of desired roots is very large. In this paper we show how exact short time dynamics can be combined with approximate frequency space results to accurately capture narrow features and a quasi-continuum of states in large molecular systems.

\begin{figure}[H]
\centering
\includegraphics[width=.9\linewidth]{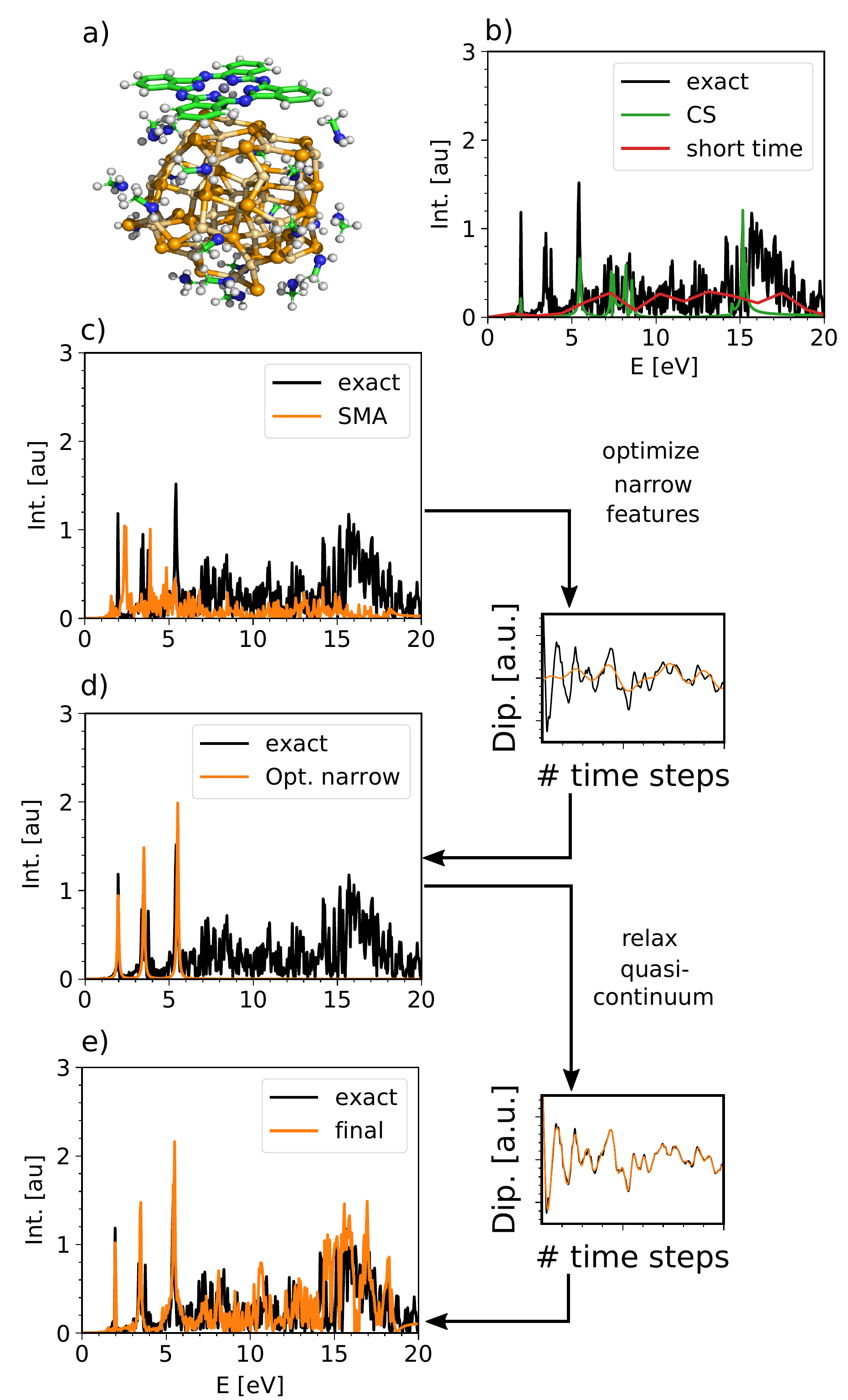}
\caption{Electronic absorption spectrum of Cd$_{38}$Se$_{38}$+ZnPc+32(NH$_2$CH$_3$), in total 301 atoms and 3,842 electrons (a). b) compares exact RT-TDDFT results (20,000 time steps) with CS and a short time simulation with both 1,000 time steps. In c), We start with an approximated spectrum which is calculated using the SMA this serves as input for d), where we use non-linear optimization to locate the narrow features. In e) we show the final spectrum after linear prediction of the quasi-continuum.}
\label{fig:intro}
\end{figure}

Our approach (BYND - Broad Yet Narrow Description) is illustrated in Figure \ref{fig:intro}, for the case of a molecular chromophore adsorbed on a surface of a semiconductor nanocrystal. In this case, a super-resolution method only captures a small number of peaks in the overall spectrum, while discrete Fourier Transform (FT) of the short time signal recovers only a broad quasi-continuum. In our approach, one first obtains an approximate spectrum - in this case using small matrix approximation (SMA) \cite{prl_sma} - that has the right number of peaks in roughly the right locations. Next, the most important narrow features in the spectrum are optimized to match the short time dynamics. Finally, linear prediction is used to match the intensities of the approximate and optimized spectral features - exactly recovering the short time signal and yielding a spectrum that is substantially more accurate than CS or discrete FT alone can provide.

BYND successfully finds electronic excitations for
\noindent large molecular systems where CS and other algorithms fail due to the presence of a quasi-continuum. For our largest test systems, we see standard mean errors between 0.01 - 0.14 eV in narrow feature position with respect to reference long-time RT-TDDFT. Considering the typical error of TDDFT with respect to experiment is around 0.25 eV \cite{doi:10.1021/ct900298e}, our method yields high quality results consistent with standard theoretical practice, useful in interpreting experimental results. Further, we see a reduction in the required computational time between 20- and 40-fold compared to standard FT due to the smaller number of time steps required by BYND. Thus, BYND enables the simulation of large molecular systems which otherwise would be computationally prohibitive even on modern computer hardware.

This article is structured as follows. We first briefly introduce the theory behind frequency-resolved approximations and exact short-time dynamics. We then move forward to a step-by-step explanation of the working equations of our method. We discuss the performance of BYND on a challenging set of large systems and conclude by discussing future directions for the method.
\section*{Results and Discussion}
\subsection*{Linear Prediction}
Modeling entire spectra from time dependent signals can, in principle, be achieved by linear prediction.\cite{doi:10.1021/cr00007a007,KOEHL1999257,Swagel:21} The basic idea is that one can predict spectral features from linear combinations of past output values. One simply determines all relevant model parameters  directly from the short-time signal.\cite{Li2020} There are techniques which achieve this in time or frequency domain and in principle, if the number of samples is sufficient and if the distance between time steps is adequate, linear prediction is able to model the spectrum with good accuracy.\cite{1162685,1451722} However, for the system sizes we are aiming for, sampling enough time steps is computationally prohibitive. Further, if we want to model a spectrum, using linear prediction only, one usually needs an idea of how many frequencies there are and where they are located.\cite{1162685,1451722} Even if we would have this information available, the number of frequencies usually exceeds the number of data points by a a large amount resulting in a under-determined system which makes it nearly impossible to extract meaningful spectra. We discuss these problems in more detail later on in this article \ref{sec:combingin_sma_with_rttddft} where we also provide examples.
\begin{figure}[h]
\centering
\includegraphics[width=1.\linewidth]{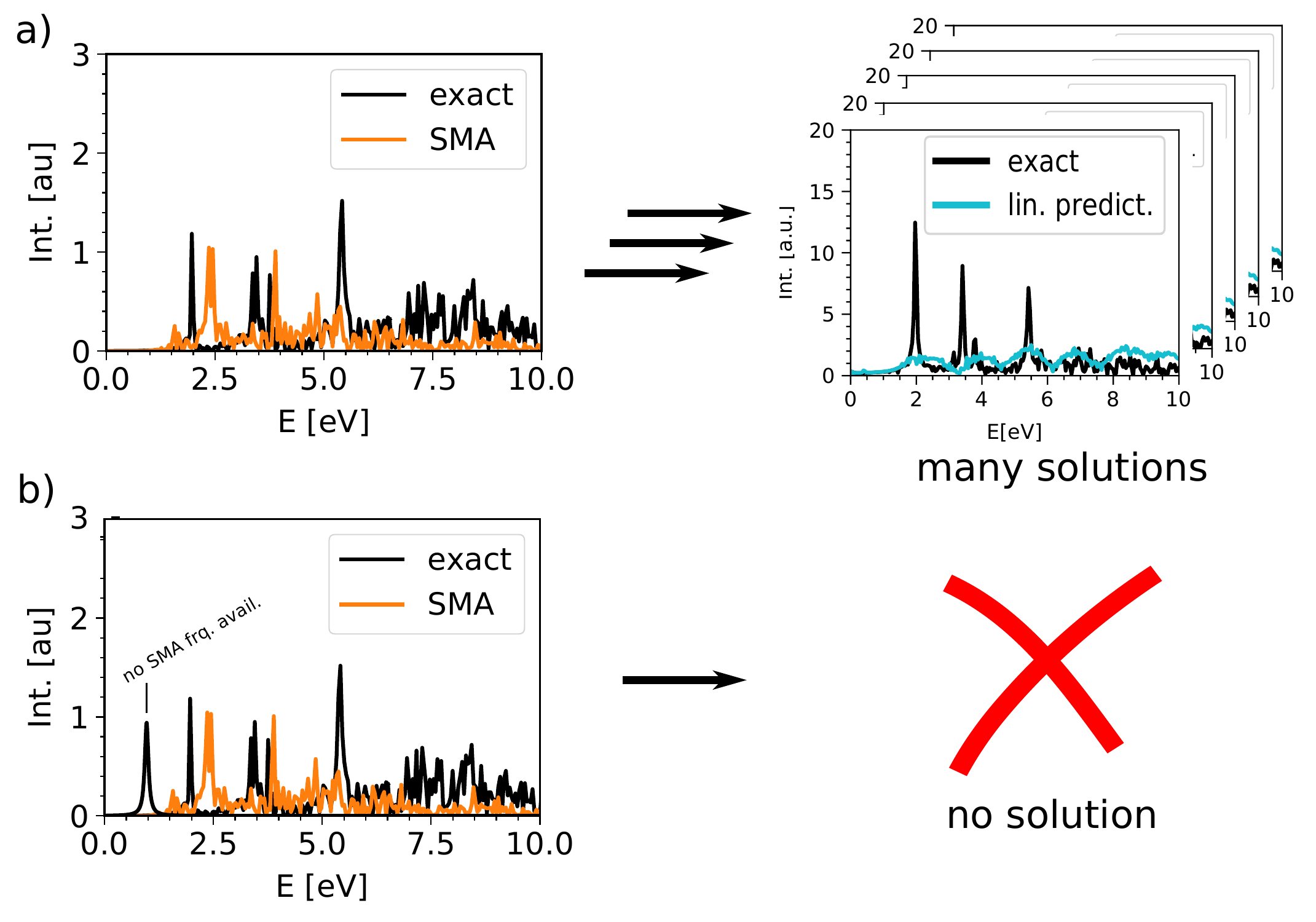}
\caption{Attempt to model the spectrum (a) of Cd$_{38}$Se$_{38}$-ZnPc-32(NH$_2$CH$_3$) with SMA frequencies and linear prediction. The short-time signal is accurately reproduced in each case, however, narrow features are completely absent in the resulting spectrum. We used a time signal with 1,000 time steps. To determine the model parameters (amplitudes) for the SMA frequencies we make use of equation \ref{eqn:ridge_regression}. In (b) we show an artificial generated spectrum where the first black exact result has no SMA frequency at the corresponding energy. In this case, linear prediction is not able to capture this feature.}
\label{fig:linear_prediction}
\end{figure}
\subsection*{SMA}
The required input for BYND is an approximated excited state spectrum which shows the right number of peaks at approximately the right energies. Therefore, we are solving the pseudo eigenvalue problem of the Casida equations\cite{casida_annual} by employing the SMA. In this approximation the electronic excitation energies in frequency space are simply given by
\begin{align}
       \omega_i = \sqrt{ \left(\epsilon_{a} - \epsilon_{i}\right)^2 + 4  \left(\epsilon_{a} - \epsilon_{i}\right) \braket{ia|f_{Hxc}|ia}}\quad,\label{eqn:sma_excitation_energies} 
\end{align}
with $\epsilon_a$ and $\epsilon_i$ being the eigenstate energies of the $a$-th virtual and $i$-th occupied Kohn-Sham state while $f_{Hxc}$ denotes the Hartree and exchange-correlation kernel. Thus the SMA gives us a simple analytical expression for obtaining a large number of excitation energies without the necessity of solving the extremely costly pseudo eigenvalue problem. We implemented the SMA within the FHIaims infrastructure. This implementation allows the rapid evaluation of several thousand exited states within systems containing easily more than 1,000 atoms (to be discussed elsewhere).% Strictly speaking, the SMA is only exact if the single-particle excitations show vanishing overlap and thus in realistic systems the SMA is error prone and can only serve as a first approximate step. However, it already contains information about the location of the "bright states".
\subsection*{RT-TDDFT}
%Strictly speaking, the SMA is only exact if the single-particle excitations show vanishing overlap and thus in realistic systems the SMA is error prone and can only serve as a first approximate step. However, it already contains information about the location of the "bright states".  
Our goal is to improve the frequency information of the SMA by combining it with exact short time dynamics from a real-time TDDFT (RT-TDDFT) simulation. In RT-TDDFT, the time-dependent Kohn-Sham states are explicitly propagated in time under the influence of an electric field which usually has the form of a sharp $\delta$-pulse, $E_{\lambda}\left(t\right)=V_{\lambda}\delta\left( t\right)$.\cite{doi:10.1021/ct500763y,Rodrigues} The effect of the electric field pulse is the excitation of all possible electronic excitation modes. Thus, the oscillation of the time-dependent dipole moment $\mu_{\nu}\left(t\right)$ from the real-time propagation can be directly linked to the excitation energies of the system. Once $\mu_{\nu}\left(t\right)$ is obtained from the simulation one can use the polarisability tensor in frequency space,\cite{ptddft,10.1063/5.0066753}
\begin{align}
       \alpha_{\lambda\nu}\left(\omega\right)=\frac{1}{V_\lambda}\int_0^{\infty}dte^{-i\omega t}\left[\mu_{\nu}\left(t\right)-\mu_{\nu}\left(t_0\right)\right]\quad, \label{eqn:polarisability_tensor} 
\end{align}
to calculate the final excitation spectrum\cite{10.1063/5.0066753}
\begin{align}
       S\left(\omega\right)= \frac{2\omega}{3\pi}\mathrm{Tr} \left\{ \Im\left[\alpha\left(\omega\right) \right]\right\} \quad. \label{eqn:osci_stength}
\end{align}
Note, that $\lambda$ and $\nu$ indicate the direction of the electric field and observed time-dependent dipole moment respectively. For a full optical excitation spectrum one needs to perform three propagations with different orientations of $E_{\lambda}$ (x, y and z).

It should be emphasized that BYND can be used with any real-time propagation method. However, due to its superior scaling with respect to system size, real-time TDDFT is the clear choice over other electronic structure methods as we attempt to push toward larger systems. In fact, real-time TDDFT is already widely employed  to capture electron dynamics in intermediately-sized molecular and solid-state systems.\cite{doi:10.1126/science.1249771,PhysRevLett.113.087401,doi:10.1021/nl100442e,https://doi.org/10.1002/adts.201800055,https://doi.org/10.1002/qua.25096,doi:10.1021/ct200137z,10.1063/5.0057587}
%Frequency space related quantities can be in principle easily extracted by standard signal processing techniques, however, these techniques often require long propagation times or fail due to the presence of a large quasi-continuum of excitations. At this point it is important to realize, that short time propagation in principle already encodes all excited state related quantities and as we later show it is already sufficient enough for obtaining accurate electronic excitations if short time dynamics is combined with the results from SMA.
\subsection*{Combining SMA with RT-TDDFT}\label{sec:combingin_sma_with_rttddft}
In this section we will introduce how our method is able to capture both narrow features and quasi-continuum by combining approximate frequency results from SMA with short time RT-TDDFT data. In order to illustrate each step of our approach the excitation spectrum of Cd$_{38}$Se$_{38}$-ZnPc-32(NH$_2$CH$_3$) will serve as a prototype (Figure \ref{fig:intro}). Within this system bulk, surface and molecular states can easily mix and the excited states of the system blur into a quasi-continuum, requiring the evaluation of a large number of excited states in even a very small energy window. In our particular example this would mean one needs to evaluate roughly 47,000 excited states in order to calculate the spectrum up to an excitation energy of 10 eV. With 3,482 total electrons, this system is highly challenging for standard TDDFT and is thus an ideal test of our approach (Figure \ref{fig:signal_dipole}). In order to extend this QD into a test set of signals representative of a broad range of common large systems, we tune its spectrum through the addition of aromatic molecules, which add narrow features, and through increasing the size of the QD, which enhances the continuum region. For our reference signal we will use short-time dynamics data containing 1,000 time steps. As we show in Figure \ref{fig:intro}a this signal length is far too short for techniques like Fourier analysis or CS to give any meaningful results. In fact, standard Fourier transform requires the simulation of 20,000 time steps in order to yield the desired resolution. 

Attempts to model the spectral features with the frequencies from SMA in a linear prediction fashion utterly fail due to the number of excitations far exceeding the number of available data points. As a consequence meaningful model parameters (amplitudes) are not extractable even by applying regularization techniques. In other cases the model frequencies from SMA might be in the wrong place and linear prediction alone even with sufficient data points is unable to extract all information. We illustrate these two cases in Figure \ref{fig:linear_prediction}. In a), the short-time signal is accurately reproduced, however, the parameters are under-determind and there are many ways to reproduce the signal. It is not possible to select a meaningful spectrum. The "bright" states are completely absent. In b), there exists no proper solution at all as there is no SMA frequency available at the position of the first bright feature. All these considerations ultimately lead to the necessity of optimizing the SMA frequencies and to restrict their number.
\subsection*{Narrow feature selection}
Our method is based on the realization that the spectrum of large systems can be separated in a sparse part and continuum part. %Methods for sparse signal recovery like CS, however, fail to locate the narrow features due to the large amount of continuum excitations and are thus not suitable for sparse feature extraction of RT-TDDFT dipole oscillations. [!!!Add here motivation why we can not use noise reduction techniques!!!]
It is important to realize that the SMA is accurate enough to  give us an estimate of how many narrow features should be present, where the narrow features are located (up to 0.5 eV accuracy), how many continuum states are present and in which frequency range the continuum is. Therefore, our decisive step is to use the SMA as an initial guess (Figure \ref{fig:signal_dipole}a). %The remaining task is then to optimize the position of the narrow features by minimizing the error with respect to the target signal. As it turns out, this optimization problem is much easier to solve if the number of narrow features is already known and can be straightforwardly achieved by a simple line search around the initial guess frequencies.
We select the initial set of narrow features by selecting each frequency for which the SMA transition dipole moment is above a certain threshold. The threshold needs to be chosen according to ensure that only "bright" excitations are included. In our example we use a threshold of 1.5 au for the intensity (Figure \ref{fig:signal_dipole}b).
\begin{algorithm*}[h]
    \caption{Line-search}\label{pseudo_code:line-search}
        \begin{algorithmic}[1]
            %\Procedure{find optimal set of frequencies}{}
            \State initial guess for $\omega_k$ from SMA
            \While {not converged}
            \State $A^{\lambda\mu}_k \gets \mathrm{min} \, \frac{1}{n} \sum_i ||y^{\lambda\mu}_i - f^{\mathrm{sparse}}_{\lambda\mu}(A^{\lambda\mu}_k,\omega_k,t_i)||^2_2 + \alpha||f^{\mathrm{sparse}}||^2_2$
            \If {iteration = 1 }
            \State $\Delta\omega \gets \Delta\omega_{\mathrm{init}}$
            \State randomly modify $A^{\lambda\mu}_k$
            \EndIf
            \For {$\omega_i\in \left\{\omega_1, ..., \omega_k\right\}$}
                \For {$\omega \in \left\{-\Delta\omega, ...,\omega_i , ...,+\Delta\omega\right\}$}
                    \State $f_{\lambda\mu} \gets -\sum_{k\neq i} A^{\lambda\mu}_{k} sin\left(\omega_k t_i\right)+A^{\lambda\mu}_i sin\left(\omega t_i\right)$
                    \State Compute $L\left(A^{\lambda\mu}_k, A^{\lambda\mu}_{i}, \omega_k,\omega\right),\quad k \neq i$
                \EndFor
                \State $\omega_i \gets \mathrm{min}\left(L\right)$
                \State $A^{\lambda\mu}_k \gets \mathrm{min} \, \frac{1}{n} \sum_i ||y^{\lambda\mu}_i - f^{\mathrm{sparse}}_{\lambda\mu}(A^{\lambda\mu}_k,\omega_k,t_i)||^2_2 + \alpha||f^{\mathrm{sparse}}||^2_2$
            \EndFor

            \EndWhile
        \end{algorithmic}
\end{algorithm*}
\subsection*{Narrow feature optimization}
The task of finding a set of optimal frequencies $\omega_k$ translates to  finding a signal $f^{\mathrm{sparse}}$ which minimizes the error with respect to the short time dynamics dipole target signal $y$. For this purpose, $f^{\mathrm{sparse}}$ at a certain time step $t_i$, can be defined as
\begin{align}
    f^{\mathrm{sparse}}_{\lambda\mu}(A^{\lambda\mu}_k,\omega_k,t_i) = -\sum_k A^{\lambda\mu}_k sin\left(\omega_k t_i\right) \quad,
\end{align}
where we make use of the fact that all excitation modes start with an in-phase oscillation right after a sharp $\delta$-pulse.\cite{doi:10.1021/acs.jctc.7b01013} Note, for cases where the electric field is parallel with the dipole operator, we are able to employ a non-negative constraint on the amplitudes.

The first step of our algorithm is to determine the amplitudes $A^{\lambda\mu}_k$ of our target frequencies $\omega_k$.  For this purpose we make use of ridge regression\cite{doi:10.1080/00401706.1970.10488634} also known as Tikhonov regularization\cite{Tikhonov:1963},
\begin{multline}\label{eqn:ridge_regression}
    \mathrm{min} \, \frac{1}{n} \sum_i ||y^{\lambda\mu}_i - f^{\mathrm{sparse}}_{\lambda\mu}(A^{\lambda\mu}_k,\omega_k,t_i)||^2_2 \\+ \alpha||f^{\mathrm{sparse}}||^2_2 \quad .
\end{multline}
Here, $\alpha$ is the regularization coefficient. Note that, in contrast to methods like CS, we do not need to enforce sparsity here as our SMA initial guess provides us with a good approximation of how many narrow features should be present. In principle, one could use CS or MUSIC for sparse feature extraction, however, we find that direct optimization of SMA provides more accurate results (see supplementary information).

Finding the optimal frequencies $\omega_k$ is a non-linear optimization problem\cite{lange2021fourier} and can be solved efficiently by performing a line-search around the initial guess for these frequencies. Our algorithm aims to minimize the following objective function,
\begin{multline}
    L\left(A^{\lambda\mu}_k,\omega_k\right) = \sum_{\lambda\mu}\sum_i ||y^{\lambda\mu}_i - f^{\mathrm{sparse}}_{\lambda\mu}(A^{\lambda\mu}_k,\omega_k,t_i)||^2_2  \\+ \beta \sum_i  A^{\lambda\mu}_k ||sin\left(\omega_k t_i\right)-sin\left(\omega^{\mathrm{init}}_k t_i\right)||^2_2 \quad ,
\end{multline}
where the first term measures the error with respect to the target signal. The last term acts as a penalty on frequencies which are too far away from their initial guess $\omega^{\mathrm{init}}_k$ with $\beta$ determining the strength of the penalty. Our procedure is realized as a greedy-algorithm\cite{CURTIS2003125}, which means our algorithm starts with the frequencies which have the highest amplitude and performs a line-search with a frequency search space $\pm\Delta\omega$ around the initial frequency. If a minimum is found the algorithm updates the old value with the newly found optimum frequency value and performs an additional amplitude adjustment step. It then moves forward to the next frequency. When all frequencies have been updated we start again by finding optimum amplitudes for the new set of frequencies. Both steps, amplitude adjustment and line-search are repeated until frequencies an amplitudes are converged. The entire procedure is described in Algorithm \ref{pseudo_code:line-search}. For more details the reader is referred to our supplementary information. 
%The amplitude adjustment step ensures that frequencies which are too far away from the target frequencies are filtered out by setting their amplitude value to zero. This is a desired effect, however, in cases where the SMA provides a non-optimal initial guess, e.g. if the frequencies are too far away from the target frequencies, this can lead to an underestimation of the number of narrow features as only frequencies with non-vanishing amplitude contribute to the objective function. This can be effectively prevented by setting  the initial value of $\Delta\omega$ to a rather high value of 0.3-0.8 au ($\Delta\omega_{\mathrm{init}}$). All following iterations use then a smaller value of 0.001 au.  Further, before we enter our line-search procedure for the first time, we modify each $A^{\lambda\mu}_k$ by adjusting their values by randomly adding a value between zero and 10\% of the highest amplitude. This ensures that every frequency will be considered in the calculation of the loss function. The entire procedure is described in Algorithm \ref{pseudo_code:line-search}.

\begin{figure}[t]
\centering
\includegraphics[width=.90\linewidth]{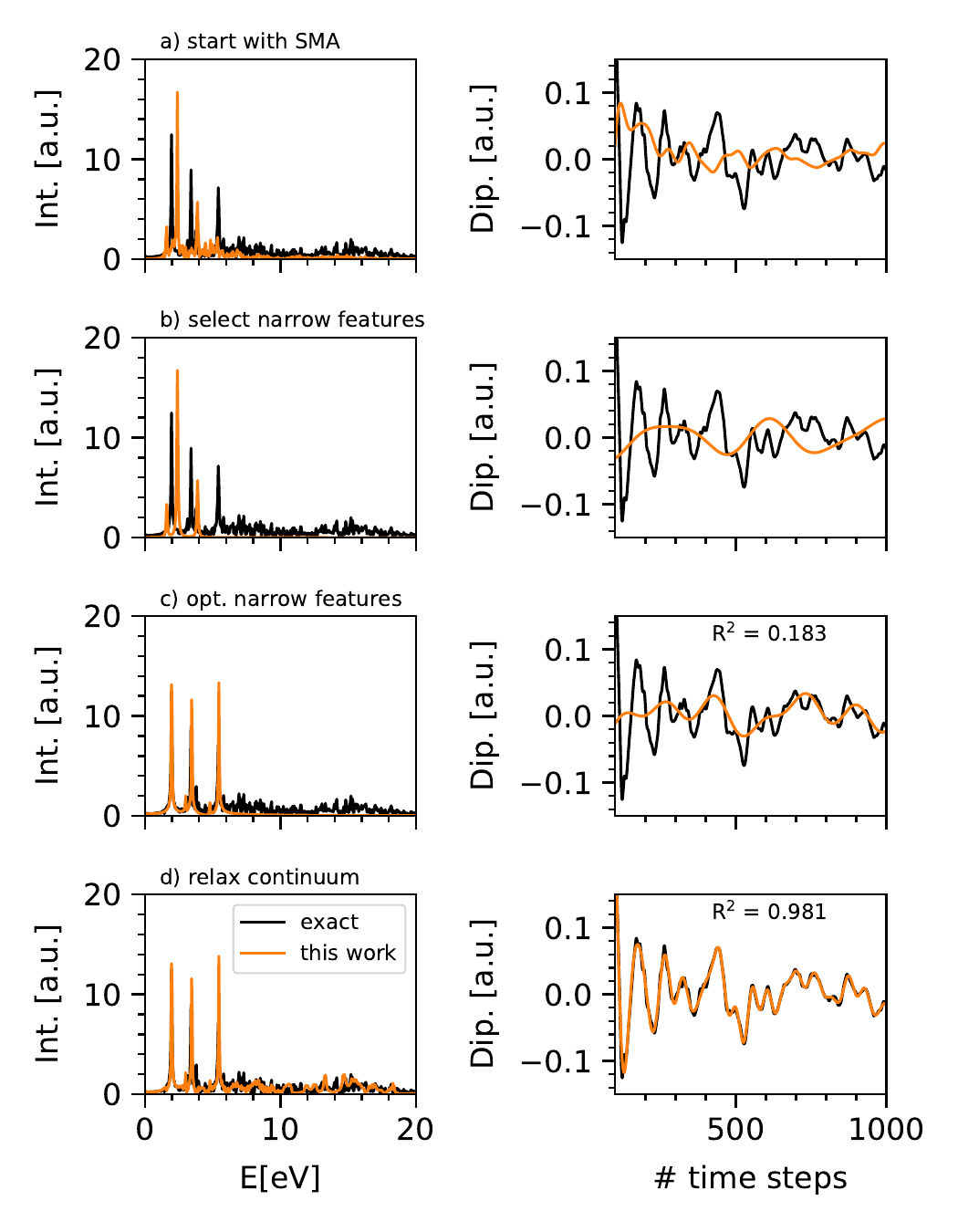}
\caption{On the left we show the Fourier transform of the time-dependent dipole moment of Cd$_{38}$Se$_{38}$-ZnPc-32(NH$_2$CH$_3$). For the sake of simplicity and without loss of generality we only shown the dipole moment in  x-direction after an electric field pulse in the same direction. On the right we show the corresponding time-dependent dipole signal. In a) we display the dipole spectrum from SMA. Panels in b) show the selected narrow features from the SMA calculations. Narrow features position after optimization are shown in c). The error with respect to exact short-time dynamics is notably reduced. The full dipole spectrum is shown in d). The full time-dependent signal is now reproduced with high accuracy.  Exact TDDFT results have been obtained from a RT-TDDFT propagation with 20,000 time steps.}
\label{fig:signal_dipole}
\end{figure}

As one can see from Figure \ref{fig:signal_dipole}c our procedure is able to successfully recover the narrow features in Cd$_{38}$Se$_{38}$-ZnPc-32(NH$_2$CH$_3$). It should be emphasized that there are in principle infinite sets of frequencies which minimise the objective function $L$. By starting with an initial guess for the number of frequencies we dramatically reduce the amount of possible solutions. Only by doing so we are able to locate the correct position of narrow features within the quasi-continuum of excitations. For the demonstration of our approach and for sake of simplicity we set $\beta$ to zero in all our test scenarios. We also note that there is the possibility to start only with signal components where $\lambda=\mu$ we observe that this can improve convergence behaviour duo to more constraints on the feature space as only non-negative amplitudes are allowed.
\subsection*{Relaxation of quasi continuum}
After optimization of the narrow features, we calculate the residual between the target signal and $f^\mathrm{sparse}$,
\begin{align}\label{eqn:residual_signal}
    y_{\lambda\mu}^{\mathrm{cont}}\left(t\right) = y_{\lambda\mu}\left(t\right)-f_{\lambda\mu}^\mathrm{sparse}\left(t\right)\quad .
\end{align}
By subtracting $f^\mathrm{sparse}$ from our target, $y^{\mathrm{cont}}$ contains only information about the continuum region of the spectrum. We now make use of the fact that the SMA contains also information about the spectral density of the continuum region and perform an additional regression in order to obtain the correct amplitudes for the continuum,
\begin{multline}
    \mathrm{min} \, \frac{1}{n} \sum_i ||y^{\mathrm{cont},\lambda\mu}_i - f^{\mathrm{cont}}_{\lambda\mu}(A^{\lambda\mu}_k,\omega_k,t_i)||^2_2 \\+ \alpha||f^{\mathrm{cont}}||^2_2 \quad .
\end{multline}
Note that in this linear prediction the index $k$ indicates the frequencies obtained from the SMA. As the target signal does not contain any narrow feature components we set the regularization coefficient $\alpha$ to the default value of 100. 
Our final reconstructed dipole signal is then given by
\begin{align}
    f_{\lambda\mu}\left(t\right) = f^\mathrm{cont}_{\lambda\mu}\left(t\right)+f_{\lambda\mu}^\mathrm{sparse}\left(t\right)\quad .
\end{align}
As we show in Figure \ref{fig:signal_dipole}d our algorithm is able to accurately reproduce the exact short time dynamics signal. We obtain amplitudes and frequencies of the "bright" states as well as correct amplitudes for the continuum region. We would like to highlight that our algorithm is completely independent of the underlying electronic structure code and can be realised in a python implementation which easily runs on standard local desktop and laptop computers.
\subsection*{Convergence and Accuracy}
\begin{figure*}[h]
\centering
\includegraphics[width=.8\linewidth]{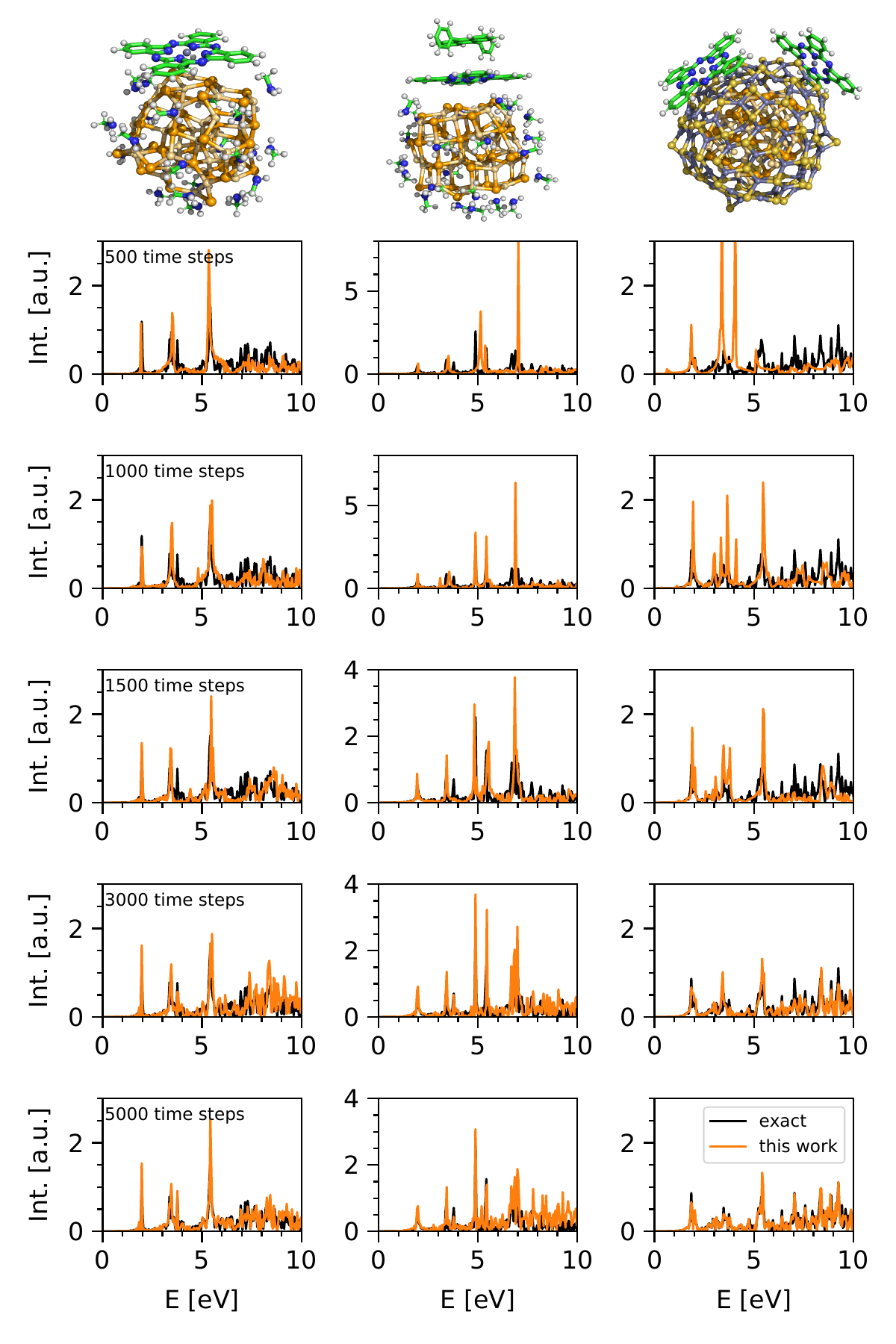}
\caption{Convergence behaviour with respect to the number of data points for the target short-time dynamics signal. We use Cd$_{38}$Se$_{38}$-ZnPc-32(NH$_2$CH$_3$) (left), Cd$_{38}$Se$_{38}$-ZnPc-DPA-32(NH$_2$CH$_3$) (middle) and Cd$_{33}$Se$_{33}$/Zn$_{93}$S$_{93}$-2(ZnPc) (right)  as a prototypical examples. We show the result for the absorption spectrum by varying the length of the short-time dynamics dipole signals between 500 and 5,000 time steps. The reference RT-TDDFT absorption spectrum was simulated with in total 20,000 time steps. For details regarding the input frequencies, the reader is referred to our supplementary information.}
\label{fig:convergence}
\end{figure*}

% explain each system why it is so challenging then explain the algorythm
Figure \ref{fig:convergence} shows the convergence of the calculated absorption spectrum with respect to the number of time steps of the target electronic dipole signals for three different systems Cd$_{38}$Se$_{38}$-ZnPc-32(NH$_2$CH$_3$), Cd$_{38}$Se$_{38}$-ZnPc-DPA-32(NH$_2$CH$_3$) and Cd$_{33}$Se$_{33}$/Zn$_{93}$S$_{93}$-2(ZnPc). These systems  demonstrate how our method performs with different types of spectra and signals. %\clearpage
%\clearpage
\noindent For Cd$_{38}$Se$_{38}$-ZnPc-DPA-32(NH$_2$CH$_3$) we expect the emergence of additional narrow features due to the presence of the DPA molecule on top of ZnPc (Figure \ref{fig:convergence}). On the contrary, the Cd$_{33}$Se$_{33}$/Zn$_{93}$S$_{93}$-2(ZnPc) system has two ZnPC molecules and a significantly larger QD size. This larger QD leads to more blurring of the "bright", localized excitations into the continuum.\afterpage{\clearpage}
\afterpage{\clearpage}
\noindent In addition, the two ZnPc molecules mimic a higher surface coverage and are on top bound to two different facets of the QD. Overall this system consists of 7,572 electrons and is thus roughly two times larger than the other two QDs and thus can be regarded as a highly challenging test case for our method.

We have additionally demonstrated the accuracy of BYND for purely molecular systems, for which the interested reader is referred to our supplementary information.
\subsection*{Observations and Trends}
As one can see in Figure \ref{fig:convergence}, we begin to obtain relatively accurate spectra compared to RT-TDDFT for our simplest system starting at only 500 time steps. Generally, all narrow features are reproduced for all test systems. Unsurprisingly, more challenging systems, namely Cd$_{33}$Se$_{33}$/Zn$_{93}$S$_{93}$-2(ZnPc), require more data points to achieve high accuracy results. However, we note that even for this system the spectrum is well reproduced using only 1,500 time steps.

In cases where the splitting between the "bright" features is small, BYND requires  more short-time data to fully resolve these details.  For example, for the "bright" states at an excitation energy of around 3.5 eV in the Cd\texorpdfstring{\textsubscript{38}}{38}Se\texorpdfstring{\textsubscript{38}}{38}-ZnPc-32(NH\texorpdfstring{\textsubscript{2}}{2}CH\texorpdfstring{\textsubscript{3}}{3}) system, our algorithm predicts one single highly "bright" feature instead of the two exact, less "bright" excitations. With such a small number of data points our algorithm is not able to distinguish between these two frequencies and more time steps are needed to resolve them. We first observe the emergence of the second "bright" excitation upon including 3,000 time steps, which is still roughly seven times shorter than the signal required for standard Fourier analysis. This is a general trend, and we obtain detailed resolved narrow features  for all test systems when using 3,000 time steps.

Similar observations can be made regarding the relative intensities. The narrow features can be clearly distinguished from the quasi-continuum background for all lengths of the short-time signal, but finding the correct relative intensities requires more time steps. The correct relative amplitudes of the narrow features are reproduced using 3,000 time steps for all considered systems. The improvement coincides with the correct assignment of continuum amplitudes. Continuum amplitudes are obtained from the residual signal $y_{\lambda\mu}^{\mathrm{cont}}$ which is described in eq. \ref{eqn:residual_signal}. As $f_{\lambda\mu}^\mathrm{sparse}$ becomes more and more accurate $y_{\lambda\mu}^{\mathrm{cont}}$ will be as well. On the other hand, an overestimation of amplitudes for "bright" excitations thus naturally leads to underestimation of the continuum region.

Going from 500 to 5,000 time steps shows a clear convergence behaviour in accuracy. At 5,000 time steps we achieve excellent agreement with the exact result which is still four times less data points compared to the full RT-TDDFT run. The fact that BYND shows a clear convergence towards the exact results underlines that BYND is not an approximate method. Provided with enough data points, BYND will yield the exact time dynamics.

Overall, our analysis shows that BYND is able to correctly predict the full excitation spectrum of large systems. Narrow features embedded in the continuum are clearly evident; bright molecular features, CT features, and contributions from the nanocrystal are all well reproduced. This is achieved while significantly reducing the required computational workload.  The range of 500 to 1,500 time steps corresponds to a speed up by a factor of 13 to 40 compared to high resolution results. For reference, this cost reduction brings the calculation of a full TDDFT spectrum for a large system close to the computational cost of a standard ground state geometry optimization.
\subsection*{Timing, Speed Up and System Size}
\begin{figure}[H]
\centering
\includegraphics[width=.9\linewidth]{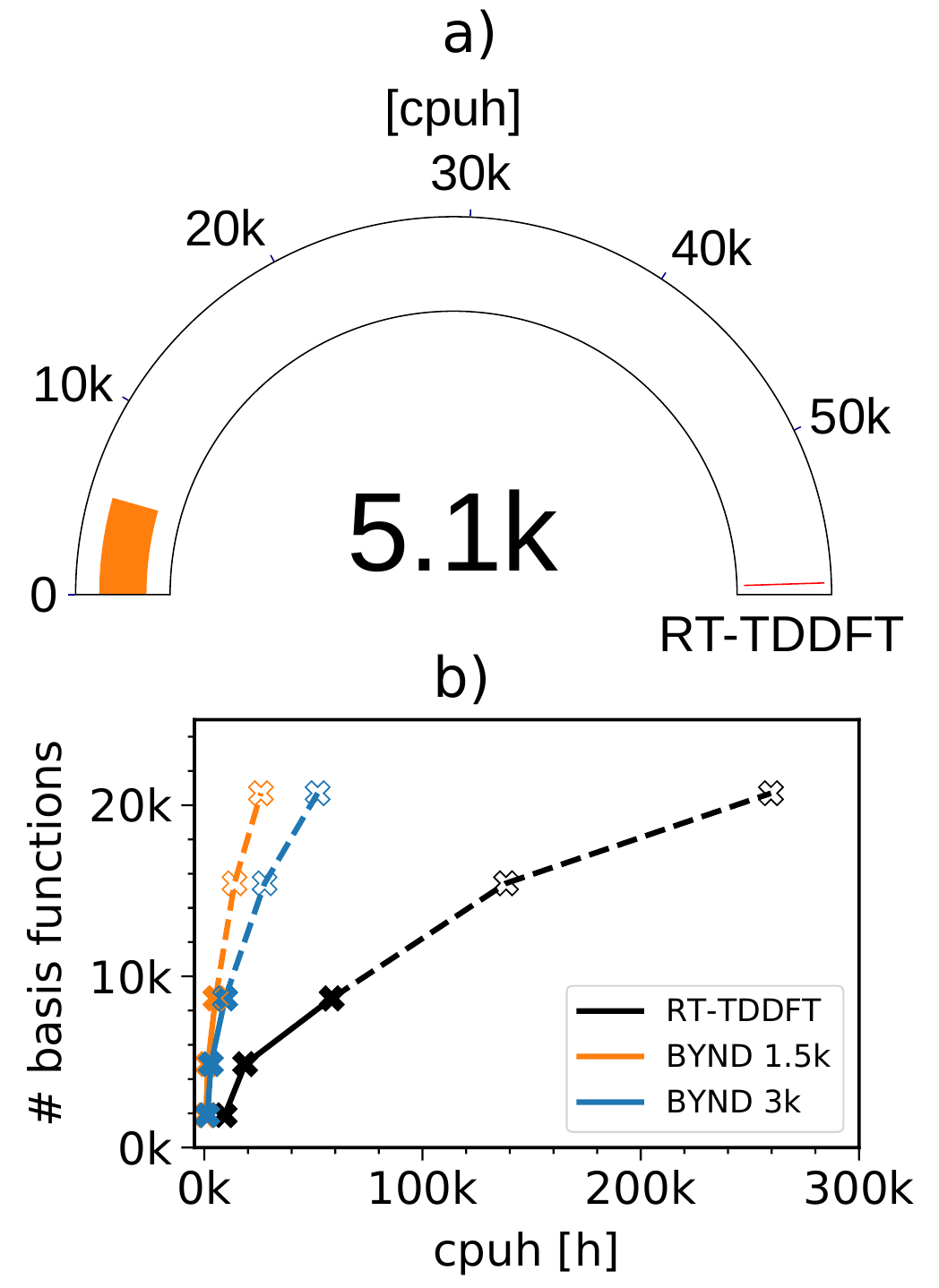}
\caption{a) CPU time required for achieving high resolution RT-TDDFT results (20,000 time steps) compared to BYND. The required time is calculated from the SMA run in combination with the required RT-TDDFT short time dynamics simulations consisting of 1,500 time steps for each direction of the applied electric field pulse. The required time for the full long time dynamics RT-TDDFT run is indicated by the red threshold on the right. We see a reduction of computational time by a factor of 11. b) We display the possible system size which can be realized with respect to the associated computational cost. We show the required time for BYND for different time steps 1,500 and 3,000 respectively.  Calculations have been performed on Intel(R) Xeon(R) Silver 4210R CPUs @ 2.40GHz.}
\label{fig:gauge}
\end{figure}

To demonstrate the speed up of our approach compared to standard high resolution long-time dynamics, we consider our largest system, Cd$_{33}$Se$_{33}$/Zn$_{93}$S$_{93}$-2(ZnPc). As we already demonstrated, for Cd$_{33}$Se$_{33}$/Zn$_{93}$S$_{93}$-2(ZnPc) we already achieve very good accuracy compared to long-time dynamics at 1,500 time steps. Thus, the total time of our approach consists of the time required for the SMA plus the time for the three RT-TDDFT short time dynamics runs with 1,500 time steps each. For the high resolution long-time dynamics simulation, the total required computational time is the cost for three RT-TDDFT runs with 20,000 time steps each. All calculations have been performed on Intel(R) Xeon(R) Silver 4210R CPUs @ 2.40GHz. Note, that we do not consider the time for our BYND itself as it runs easily on a standard laptop computer and usually converges within several minutes only. Thus, adding only a vanishing amount of computational time. In total, BYND needs 4,679 CPUh compared to 52,234 CPUh for the full long-time dynamics run which translates into a reduction of required computational time by factor of 11.2 (Figure \ref{fig:gauge}a). Another perspective is to look at the required time for a given system size. Figure \ref{fig:gauge}b shows the system size which is possible at a given computational cost. The filled data points represent f-cororene, Cd$_{38}$Se$_{38}$-ZnPc-32(NH$_2$CH$_3$) and Cd$_{33}$Se$_{33}$/Zn$_{93}$S$_{93}$-2(ZnPc) which have been fully simulated. Unfilled marker represent additional nanocrystal systems (see supplementary information) where we used the RT-TDDFT walltime estimate provided by FHIaims in order to predict the computational cost. At a given walltime, BYND enables the simulation of significantly larger systems compared to a standard RT-TDDFT run. Even with an input signal of 3,000 time steps, BYND outperforms standard RT-TDDFT by far regarding possible system size.
%\subsection*{Conclusion \& Outlook}
%\section*{Discussion}
In this work we showed how approximate frequency space results can be combined with short time dynamics simulations in order to accurately capture narrow features and a quasi-continuum of states for large systems. Due to the ability of BYND to use only short time dynamics, we are able to significantly reduce the computational time which is needed for the underlying electronic structure simulations. For one of our highly challenging systems we observe a reduction by a factor of 11.  The reduction of computational time is due to two key components of our approach. First, we use the SMA as an estimate for how many narrow features can be expected and which frequencies they have. Second, we use this information to further optimize their position and amplitudes by minimizing the error with respect to the short-time dynamics signal which on its own would have insignificant resolution to capture the spectrum. Thus, our approach allows researchers to understand the electronic properties of large systems which were previously computationally inaccessible.
In contrast to methods such as  filter diagonalization\cite{10.1063/1.468999}, which only shows promising results if the spectrum is not too dense\cite{doi:10.1021/acs.jctc.6b00511}, BYND is explicitly designed to work with high spectral densities. Further, we would like to emphasise that BYND is not an approximation: if enough data is provided, the results will always converge towards the full time dynamics of the chosen electronic structure method. This is in contrast to other methods such as simplified TDDFT\cite{BANNWARTH201445}, simplified GW/sBSE\cite{doi:10.1021/acs.jctc.2c00087}, or TD-INDO/S\cite{doi:10.1021/acs.jctc.7b00618} which employ approximations to the electron interaction integrals in order to achieve computational speedup. To increase the data available without computing more time steps, future work on BYND is aimed towards including quadrupole or higher multipole moments. More data points will then enable the use of even shorter time dynamics. Improvement in accuracy could be achieved by using the Casida equations to explicitly describe just the first few  excited states, which can be then fixed in our non-linear optimization to yield an even more efficient localization of the remaining narrow features. Furthermore, we see potential in improving our line-search routine by employing advanced machine-learning techniques, which may allow us to further increase the range of the spectrum covered by the sparse signal. Further one can use SMA results from semi-local exchange-correlation kernels to approximate hybrid TDDFT or GW results, thus saving additional simulation time.

In conclusion, we have combined frequency domain results with exact short time dynamics in order to create a super-resolution technique (BYND) which allows for the \emph{ab initio} description of the entire excitation spectrum for systems which are beyond the system size boundaries of current electronic structure methods.
\subsection*{Methods}
All TDDFT calculations have been carried out using the FHIaims\cite{BLUM20092175} program. Exchange-correlation interactions have been treated using the PBE\cite{PhysRevLett.77.3865} functional. \emph{Light} \emph{tier1} settings have been used for the integration grid and basis set.  All RT-TDDFT\cite{10.1063/5.0066753} calculations have been performed with a time step of 0.2 au and an electric field strength of 0.01 au. Total simulation time was 4,000 au.
\subsection*{Supporting Information Appendix (SI)}
In our supporting information we discuss the initialization procedure of BYND and demonstrate the robustness of our algorithm regarding the initial guess. Further, we demonstrate that our algorithm also yields accurate results for other different system types like small to medium sized molecules. We also show a detailed comparison with CS and show the influence of the $\alpha$ regularization parameter in the ridge regression step. A comparison with the Pad\'e approximation is also included in our SI.
\subsection*{Acknowledgement}
M.K. acknowledges support from the German Research Foundation (DFG, KI 2558/1-1). Further, M.K. would like to thank Prof. Harald Oberhofer and Cristina Grosu for their valuable input and support.

\bibliographystyle{plain}
\bibliography{sample}

\begin{thebibliography}{10}

\bibitem{REN2021113087}
Refined standards for simulating uv–vis absorption spectra of acceptors in organic solar cells by td-dft.
\newblock {\em Journal of Photochemistry and Photobiology A: Chemistry}, 407:113087, 2021.

\bibitem{D0CP00060D}
Amjad Ali, Muhammad~Imran Rafiq, Zhuohan Zhang, Jinru Cao, Renyong Geng, Baojing Zhou, and Weihua Tang.
\newblock Td-dft benchmark for uv-visible spectra of fused-ring electron acceptors using global and range-separated hybrids.
\newblock {\em Phys. Chem. Chem. Phys.}, 22:7864--7874, 2020.

\bibitem{BANNWARTH201445}
Christoph Bannwarth and Stefan Grimme.
\newblock A simplified time-dependent density functional theory approach for electronic ultraviolet and circular dichroism spectra of very large molecules.
\newblock {\em Computational and Theoretical Chemistry}, 1040-1041:45--53, 2014.
\newblock Excited states: From isolated molecules to complex environments.

\bibitem{BLUM20092175}
Volker Blum, Ralf Gehrke, Felix Hanke, Paula Havu, Ville Havu, Xinguo Ren, Karsten Reuter, and Matthias Scheffler.
\newblock Ab initio molecular simulations with numeric atom-centered orbitals.
\newblock {\em Computer Physics Communications}, 180(11):2175--2196, 2009.

\bibitem{doi:10.1021/acs.jctc.6b00511}
Adam Bruner, Daniel LaMaster, and Kenneth Lopata.
\newblock Accelerated broadband spectra using transition dipole decomposition and padé approximants.
\newblock {\em Journal of Chemical Theory and Computation}, 12(8):3741--3750, 2016.
\newblock PMID: 27359347.

\bibitem{compressed_sensing}
Emmanuel~J. Candès, Justin~K. Romberg, and Terence Tao.
\newblock Stable signal recovery from incomplete and inaccurate measurements.
\newblock {\em Communications on Pure and Applied Mathematics}, 59(8):1207--1223, 2006.

\bibitem{casida_annual}
M.E. Casida and M.~Huix-Rotllant.
\newblock Progress in time-dependent density-functional theory.
\newblock {\em Annual Review of Physical Chemistry}, 63(1):287--323, 2012.
\newblock PMID: 22242728.

\bibitem{doi:10.1021/acs.jctc.2c00087}
Yeongsu Cho, Sylvia~J. Bintrim, and Timothy~C. Berkelbach.
\newblock Simplified gw/bse approach for charged and neutral excitation energies of large molecules and nanomaterials.
\newblock {\em Journal of Chemical Theory and Computation}, 18(6):3438--3446, 2022.
\newblock PMID: 35544591.

\bibitem{COPPOLA2022125603}
Carmen Coppola, Rossella Infantino, Alessio Dessì, Lorenzo Zani, Maria~Laura Parisi, Alessandro Mordini, Gianna Reginato, Riccardo Basosi, and Adalgisa Sinicropi.
\newblock Dft and tddft investigation of four triphenylamine/phenothiazine-based molecules as potential novel organic hole transport materials for perovskite solar cells.
\newblock {\em Materials Chemistry and Physics}, 278:125603, 2022.

\bibitem{CURTIS2003125}
S.A. Curtis.
\newblock The classification of greedy algorithms.
\newblock {\em Science of Computer Programming}, 49(1):125--157, 2003.

\bibitem{doi:10.1126/science.1249771}
Sarah~Maria Falke, Carlo~Andrea Rozzi, Daniele Brida, Margherita Maiuri, Michele Amato, Ephraim Sommer, Antonietta~De Sio, Angel Rubio, Giulio Cerullo, Elisa Molinari, and Christoph Lienau.
\newblock Coherent ultrafast charge transfer in an organic photovoltaic blend.
\newblock {\em Science}, 344(6187):1001--1005, 2014.

\bibitem{Gasparini2021}
Nicola Gasparini, Franco V.~A. Camargo, Stefan Fr{\"u}hwald, Tetsuhiko Nagahara, Andrej Classen, Steffen Roland, Andrew Wadsworth, Vasilis~G. Gregoriou, Christos~L. Chochos, Dieter Neher, Michael Salvador, Derya Baran, Iain McCulloch, Andreas G{\"o}rling, Larry L{\"u}er, Giulio Cerullo, and Christoph~J. Brabec.
\newblock Adjusting the energy of interfacial states in organic photovoltaics for maximum efficiency.
\newblock {\em Nature Communications}, 12(1):1772, Mar 2021.

\bibitem{doi:10.1021/acs.jctc.7b00618}
Soumen Ghosh, Amity Andersen, Laura Gagliardi, Christopher~J. Cramer, and Niranjan Govind.
\newblock Modeling optical spectra of large organic systems using real-time propagation of semiempirical effective hamiltonians.
\newblock {\em Journal of Chemical Theory and Computation}, 13(9):4410--4420, 2017.
\newblock PMID: 28813603.

\bibitem{Goldzak2021}
Tamar Goldzak, Alexandra~R. McIsaac, and Troy Van~Voorhis.
\newblock Colloidal cdse nanocrystals are inherently defective.
\newblock {\em Nature Communications}, 12(1):890, Feb 2021.

\bibitem{10.1063/5.0066753}
Joscha Hekele, Yi~Yao, Yosuke Kanai, Volker Blum, and Peter Kratzer.
\newblock {All-electron real-time and imaginary-time time-dependent density functional theory within a numeric atom-centered basis function framework}.
\newblock {\em The Journal of Chemical Physics}, 155(15):154801, 10 2021.

\bibitem{doi:10.1080/00401706.1970.10488634}
Arthur~E. Hoerl and Robert~W. Kennard.
\newblock Ridge regression: Biased estimation for nonorthogonal problems.
\newblock {\em Technometrics}, 12(1):55--67, 1970.

\bibitem{doi:10.1021/ct900298e}
Denis Jacquemin, Valérie Wathelet, Eric~A. Perpète, and Carlo Adamo.
\newblock Extensive td-dft benchmark: Singlet-excited states of organic molecules.
\newblock {\em Journal of Chemical Theory and Computation}, 5(9):2420--2435, 2009.
\newblock PMID: 26616623.

\bibitem{ptddft}
Joaquim Jornet-Somoza and Irina Lebedeva.
\newblock Real-time propagation tddft and density analysis for exciton coupling calculations in large systems.
\newblock {\em Journal of Chemical Theory and Computation}, 15(6):3743--3754, 2019.
\newblock PMID: 31091099.

\bibitem{Kinoshita2015}
Takumi Kinoshita, Kazuteru Nonomura, Nam Joong~Jeon, Fabrizio Giordano, Antonio Abate, Satoshi Uchida, Takaya Kubo, Sang~Il Seok, Mohammad~Khaja Nazeeruddin, Anders Hagfeldt, Michael Gr{\"a}tzel, and Hiroshi Segawa.
\newblock Spectral splitting photovoltaics using perovskite and wideband dye-sensitized solar cells.
\newblock {\em Nature Communications}, 6(1):8834, Nov 2015.

\bibitem{KOEHL1999257}
P.~Koehl.
\newblock Linear prediction spectral analysis of nmr data.
\newblock {\em Progress in Nuclear Magnetic Resonance Spectroscopy}, 34(3):257--299, 1999.

\bibitem{lange2021fourier}
Henning Lange, Steven~L Brunton, and J~Nathan Kutz.
\newblock From fourier to koopman: Spectral methods for long-term time series prediction.
\newblock {\em The Journal of Machine Learning Research}, 22(1):1881--1918, 2021.

\bibitem{doi:10.1021/cr00007a007}
Jens~J. Led and Henrik. Gesmar.
\newblock Application of the linear prediction method to nmr spectroscopy.
\newblock {\em Chemical Reviews}, 91(7):1413--1426, 1991.

\bibitem{Li2020}
Ruixiang Li, Hui Li, and Weibin Shi.
\newblock Human activity recognition based on lpa.
\newblock {\em Multimedia Tools and Applications}, 79(41):31069--31086, Nov 2020.

\bibitem{https://doi.org/10.1002/adts.201800055}
Chao Lian, Mengxue Guan, Shiqi Hu, Jin Zhang, and Sheng Meng.
\newblock Photoexcitation in solids: First-principles quantum simulations by real-time tddft.
\newblock {\em Advanced Theory and Simulations}, 1(8):1800055, 2018.

\bibitem{doi:10.1021/ct200137z}
Kenneth Lopata and Niranjan Govind.
\newblock Modeling fast electron dynamics with real-time time-dependent density functional theory: Application to small molecules and chromophores.
\newblock {\em Journal of Chemical Theory and Computation}, 7(5):1344--1355, 2011.
\newblock PMID: 26610129.

\bibitem{10.1093/oxfmat/itad003}
Melchizedek Lyakurwa and Surendra~Babu Numbury.
\newblock {DFT and TD-DFT study of Optical and Electronic Properties of new donor–acceptor–donor monomers for polymer solar cells}.
\newblock {\em Oxford Open Materials Science}, 3(1):itad003, 03 2023.

\bibitem{1451722}
J.~Makhoul.
\newblock Linear prediction: A tutorial review.
\newblock {\em Proceedings of the IEEE}, 63(4):561--580, 1975.

\bibitem{1162685}
J.~Makhoul.
\newblock Spectral linear prediction: Properties and applications.
\newblock {\em IEEE Transactions on Acoustics, Speech, and Signal Processing}, 23(3):283--296, 1975.

\bibitem{258082}
S.G. Mallat and Zhifeng Zhang.
\newblock Matching pursuits with time-frequency dictionaries.
\newblock {\em IEEE Transactions on Signal Processing}, 41(12):3397--3415, 1993.

\bibitem{doi:10.1021/nl100442e}
Sheng Meng and Efthimios Kaxiras.
\newblock Electron and hole dynamics in dye-sensitized solar cells: Influencing factors and systematic trends.
\newblock {\em Nano Letters}, 10(4):1238--1247, 2010.
\newblock PMID: 20353199.

\bibitem{D2RA06880J}
Batool Moradpour and Reza Omidyan.
\newblock Dft/td-dft study of electronic and phosphorescent properties in cycloplatinated complexes: implications for oleds.
\newblock {\em RSC Adv.}, 12:34217--34225, 2022.

\bibitem{Neef2023}
Alexander Neef, Samuel Beaulieu, Sebastian Hammer, Shuo Dong, Julian Maklar, Tommaso Pincelli, R.~Patrick Xian, Martin Wolf, Laurenz Rettig, Jens Pflaum, and Ralph Ernstorfer.
\newblock Orbital-resolved observation of singlet fission.
\newblock {\em Nature}, 616(7956):275--279, Apr 2023.

\bibitem{Orovic2016}
Irena Orovi{\'{c}}, Vladan Papi{\'{c}}, Cornel Ioana, Xiumei Li, and Srdjan Stankovi{\'{c}}.
\newblock Compressive sensing in signal processing: Algorithms and transform domain formulations.
\newblock {\em Mathematical Problems in Engineering}, 2016:7616393, Oct 2016.

\bibitem{Rodrigues}
Ronaldo~Rodrigues Pela and Claudia Draxl.
\newblock All-electron full-potential implementation of real-time tddft in exciting.
\newblock {\em Electronic Structure}, 3(3):037001, jul 2021.

\bibitem{PhysRevLett.77.3865}
John~P. Perdew, Kieron Burke, and Matthias Ernzerhof.
\newblock Generalized gradient approximation made simple.
\newblock {\em Phys. Rev. Lett.}, 77:3865--3868, Oct 1996.

\bibitem{PhysRevLett.128.016801}
E.~Perfetto, Y.~Pavlyukh, and G.~Stefanucci.
\newblock Real-time $gw$: Toward an ab initio description of the ultrafast carrier and exciton dynamics in two-dimensional materials.
\newblock {\em Phys. Rev. Lett.}, 128:016801, Jan 2022.

\bibitem{https://doi.org/10.1002/qua.25096}
Makenzie~R. Provorse and Christine~M. Isborn.
\newblock Electron dynamics with real-time time-dependent density functional theory.
\newblock {\em International Journal of Quantum Chemistry}, 116(10):739--749, 2016.

\bibitem{10.1063/5.0004865}
J.~J. Rehr, F.~D. Vila, J.~J. Kas, N.~Y. Hirshberg, K.~Kowalski, and B.~Peng.
\newblock {Equation of motion coupled-cluster cumulant approach for intrinsic losses in x-ray spectra}.
\newblock {\em The Journal of Chemical Physics}, 152(17):174113, 05 2020.

\bibitem{C7CP07849H}
M.~Ruberti, P.~Decleva, and V.~Averbukh.
\newblock Multi-channel dynamics in high harmonic generation of aligned co2: ab initio analysis with time-dependent b-spline algebraic diagrammatic construction.
\newblock {\em Phys. Chem. Chem. Phys.}, 20:8311--8325, 2018.

\bibitem{doi:10.1021/acs.jctc.7b01013}
Ingo Schelter and Stephan Kümmel.
\newblock Accurate evaluation of real-time density functional theory providing access to challenging electron dynamics.
\newblock {\em Journal of Chemical Theory and Computation}, 14(4):1910--1927, 2018.

\bibitem{music}
R.~Schmidt.
\newblock Multiple emitter location and signal parameter estimation.
\newblock {\em IEEE Transactions on Antennas and Propagation}, 34(3):276--280, 1986.

\bibitem{1143830}
R.~Schmidt.
\newblock Multiple emitter location and signal parameter estimation.
\newblock {\em IEEE Transactions on Antennas and Propagation}, 34(3):276--280, 1986.

\bibitem{SEJDIC201822}
Ervin Sejdic, Irena Orovic, and Srdjan Stankovic.
\newblock Compressive sensing meets time–frequency: An overview of recent advances in time–frequency processing of sparse signals.
\newblock {\em Digital Signal Processing}, 77:22--35, 2018.
\newblock Digital Signal Processing \& SoftwareX - Joint Special Issue on Reproducible Research in Signal Processing.

\bibitem{10.1063/5.0057587}
Christopher Shepard, Ruiyi Zhou, Dillon~C. Yost, Yi~Yao, and Yosuke Kanai.
\newblock {Simulating electronic excitation and dynamics with real-time propagation approach to TDDFT within plane-wave pseudopotential formulation}.
\newblock {\em The Journal of Chemical Physics}, 155(10):100901, 09 2021.

\bibitem{SLOWIK2017132}
Irma Slowik, Axel Fischer, Hartmut Fröb, Simone Lenk, Sebastian Reineke, and Karl Leo.
\newblock Novel organic light-emitting diode design for future lasing applications.
\newblock {\em Organic Electronics}, 48:132--137, 2017.

\bibitem{PhysRev.84.244}
R.~Sternheimer.
\newblock On nuclear quadrupole moments.
\newblock {\em Phys. Rev.}, 84:244--253, Oct 1951.

\bibitem{Swagel:21}
E.~Swagel, J.~Paul, A.~D. Bristow, and J.~K. Wahlstrand.
\newblock Analysis of complex multidimensional optical spectra by linear prediction.
\newblock {\em Opt. Express}, 29(23):37525--37533, Nov 2021.

\bibitem{Tikhonov:1963}
A.~N. Tikhonov.
\newblock Solution of incorrectly formulated problems and the regularization method.
\newblock {\em Soviet Math. Dokl.}, 4:1035--1038, 1963.

\bibitem{doi:10.1021/ct500763y}
Samat Tussupbayev, Niranjan Govind, Kenneth Lopata, and Christopher~J. Cramer.
\newblock Comparison of real-time and linear-response time-dependent density functional theories for molecular chromophores ranging from sparse to high densities of states.
\newblock {\em Journal of Chemical Theory and Computation}, 11(3):1102--1109, 2015.

\bibitem{prl_sma}
Igor Vasiliev, Serdar \"O\ifmmode~\breve{g}\else \u{g}\fi{}\"ut, and James~R. Chelikowsky.
\newblock Ab initio excitation spectra and collective electronic response in atoms and clusters.
\newblock {\em Phys. Rev. Lett.}, 82:1919--1922, Mar 1999.

\bibitem{doi:10.1021/acs.jctc.0c00639}
F.~D. Vila, J.~J. Rehr, J.~J. Kas, K.~Kowalski, and B.~Peng.
\newblock Real-time coupled-cluster approach for the cumulant green’s function.
\newblock {\em Journal of Chemical Theory and Computation}, 16(11):6983--6992, 2020.
\newblock PMID: 33108872.

\bibitem{10.1063/5.0099192}
Fernando~D. Vila, John~J. Rehr, Himadri Pathak, Bo~Peng, Ajay Panyala, Erdal Mutlu, Nicholas~P. Bauman, and Karol Kowalski.
\newblock {Real-time equation-of-motion CC cumulant and CC Green’s function simulations of photoemission spectra of water and water dimer}.
\newblock {\em The Journal of Chemical Physics}, 157(4):044101, 08 2022.

\bibitem{PhysRevLett.113.087401}
Georg Wachter, Christoph Lemell, Joachim Burgd\"orfer, Shunsuke~A. Sato, Xiao-Min Tong, and Kazuhiro Yabana.
\newblock Ab initio simulation of electrical currents induced by ultrafast laser excitation of dielectric materials.
\newblock {\em Phys. Rev. Lett.}, 113:087401, Aug 2014.

\bibitem{10.1063/1.468999}
Michael~R. Wall and Daniel Neuhauser.
\newblock {Extraction, through filter‐diagonalization, of general quantum eigenvalues or classical normal mode frequencies from a small number of residues or a short‐time segment of a signal. I. Theory and application to a quantum‐dynamics model}.
\newblock {\em The Journal of Chemical Physics}, 102(20):8011--8022, 05 1995.

\bibitem{omp}
Jian Wang, Seokbeop Kwon, and Byonghyo Shim.
\newblock Generalized orthogonal matching pursuit.
\newblock {\em IEEE Transactions on Signal Processing}, 60(12):6202--6216, 2012.

\bibitem{ZAIER2019108}
Rania Zaier, Said Hajaji, Masatoshi Kozaki, and Sahbi Ayachi.
\newblock Dft and td-dft studies on the electronic and optical properties of linear $\pi$-conjugated cyclopentadithiophene (cpdt) dimer for efficient blue oled.
\newblock {\em Optical Materials}, 91:108--114, 2019.

\end{thebibliography}

\end{document}